\documentstyle[aps,epsf,multicol]{revtex}
\begin{document}
\draft
\widetext
\title{Andreev reflection in the fractional quantum Hall effect}
\author{Nancy P. Sandler, Claudio de C. Chamon and Eduardo Fradkin}
\address{Department of Physics, University of Illinois at Urbana-Champaign,
Urbana, IL 61801-3080}
\maketitle
\begin{abstract}
We study the reflection of electrons and quasiparticles on
point-contact interfaces between fractional quantum Hall (FQH) states
and normal metals (leads), as well as interfaces between two FQH
states with mismatched filling fractions. We classify the processes
taking place at the interface in the strong coupling limit. In this
regime a set of quasiparticles can decay into quasiholes on the FQH side
and charge excitations on the other side of the junction.  This
process is analogous to an Andreev reflection in
normal-metal/superconductor (N-S) interfaces.
\end{abstract}



\begin{multicols}{2}
\narrowtext
\section{INTRODUCTION}
\label{sec:intro}

The first experimental manifestation of the quantized Hall effect came in
transport measurements\cite{Klitzing}: a precisely defined
fractional Hall conductance $\sigma_{xy}=\frac{p}{q}\frac{e^2}{h}$ and
a vanishing longitudinal conductance $\sigma_{xx}$. These transport
properties can be explained by looking at the spectrum of an isolated
quantum Hall state. A gauge invariance argument, proposed by
Laughlin\cite{Laughlin} and elaborated by
Halperin\cite{Halperin}, relates the existence of a gap for
current carrying states to the fractionally quantized Hall conductance
$\sigma_{xy}=\frac{p}{q}\frac{e^2}{h}$. It is interesting to parallel
the case of quantized Hall effects to that of superconductivity, where
a transport property, a vanishing resistivity, follows from the
existence of an energy gap in the system. In both cases the study of
the isolated system, such as the spectrum and quasiparticle
excitations, provides the answers for the features observed via
transport measurements.

However, as we know for the case of superconductors, there are
interesting physical phenomena which arise not from isolated systems,
but from the contact of the system with a normal metal. For example,
an electron incident from the normal metal side can either be
back-reflected at the interface, or be Andreev reflected
\cite{Andreev} as a hole and transfer charge $2e$ (a Cooper pair
\cite{Footnote1}) to the superconductor. It is then
very natural to ask whether similar effects can also occur in the case
of the FQH effect. More precisely, we should ask what happens in FQH
junctions, {\it i.e.}, when we bring a FQH state in contact with
either a normal metal (leads) or another FQH state with different
filling fraction.

Some of the properties of these FQH junctions resemble those of
normal-metal/superconductor (N-S) junctions, even though the
underlying physical reasons are quite different. In the case of
Andreev reflection in N-S junctions, an electron incoming from the N
side with an energy falling within the superconducting gap cannot go
to the S side. It costs a finite energy to create a quasiparticle
excitation in the superconductor, but not to create a Cooper pair, so
the incoming electron from the N side can be reflected as a hole in
the N side, leaving charge $2e$ on the S side of the junction. Here
the superconducting gap plays a fundamental role. Now, in the case of
FQH junctions, there is a gap for all excitations in the bulk but
there are always gapless excitations at the boundary, the edge states.
Thus, the mechanism for reflection processes at the boundary does not
depend directly on the gap, but instead on the topological properties
of FQH states.  Consider, for example, a point-contact junction
between a $\nu=1/3$ FQH state and either leads or a $\nu=1$
state. Quasiparticles with fractional charge $e^*=e/3$ incoming from
the $\nu=1/3$ side cannot go to the leads or $\nu=1$ side, for the
states on the other side of the junction do not sustain fractionally
charged excitations (one may think of this as an infinite gap for
charge $e^*=e/3$ excitations in a normal metal or a $\nu=1$
state). Thus, upon reaching the junction, a state with two $\nu=1/3$
quasiparticles can (in addition to simply be back-reflected) be
reflected as one quasihole with charge $-e/3$ while transmitting
charge $e$ to the other side of the junction.

In this paper we will study the different reflection processes taking
place at point-contact interfaces between FQH states and normal metals
or between two FQH states with different filling fractions. The tools
for the study of this problem have been developed in Ref. \cite{CF},
where the problem of tunneling between two chiral Luttinger liquids
with different $g$ parameters was treated by mapping the problem to
tunneling between Luttinger liquids with same $g$, and exploring the
weak-strong coupling duality symmetry in the latter problem. In that
work the problem was studied at the level of the boson fields without
considering the underlying Hilbert space for scattering between
solitons representing electrons on one side, and fractionally charged
quasiparticles on the other.

In what follows we will be able to write the quasiparticle operators
before scattering in terms of the chiral boson fields describing the
edge excitations. Whereas for weak coupling the quasiparticle
operators can still be written in terms of their original chiral boson
fields, for strong coupling the boson fields describing both sides of
the junction get mixed, and the mixing is determined by the filling
fraction mismatch.  Using the scattering of quasiparticle operators at
the junction we obtain a {\it selection matrix} $M$ which relates {\it
quantum numbers} of incoming and outgoing quasiparticle or soliton
states. Such one to one correspondence between in and out quantum
numbers exists at both weak and strong coupling limits.  We will show
that at weak coupling the selection matrix is trivial but at strong
coupling it depends on the value of the filling fraction of the FQH
system.  Although this matrix contains less information than the ${\bf
S}$-matrix (which relates amplitudes), it is sufficient to determine
transport properties such as the conductance and it allows a
classification of the possible processes taking place at the junction.
The selection matrix encodes a set of selection rules which we show
can be satisfied if the quantum numbers of the scattering states lie
on a 2D lattice. At strong coupling the lattice can be described by
two basis vectors corresponding to two different reflection processes:
normal and Andreev type reflections.  The paper is organized as
follows: in Sec.~\ref{sec:model}, we describe the model for a
FQH-normal metal junction at a point contact. In
Sec.~\ref{sec:analysis} we analyze the model at both fixed points:
weak and strong. In Sec.~\ref{sec:classif} we classify the allowed
scattering processes at the strong coupling fixed point, in
Sec.~\ref{sec:kondo} we present a conjecture that relates the FQH
tunnel junction to the two-channel Kondo problem, and in
Sec.~\ref{sec:conclusions} we review our main results. Details on the
proper definition of electron operators and the role of boundary
conditions are discussed in the Appendix.

\section{Formulation of the model}
\label{sec:model}

We start with a model Lagrangian for the FQH-normal metal junction at
a point-contact which describes the dynamics on the edge of a FQH
liquid, the electron gas reservoirs, and the tunneling between them
through a single point-contact of the form
\begin{equation}
{\cal L} = {\cal L}_{edge} + {\cal L}_{res} + {\cal L}_{tun}\,.
\label{eq:Lag}
\end{equation}
The dynamics of the edge of the FQH liquid with a Laughlin filling
fraction $\nu=\frac{1}{2k+1}$ is described by a free chiral boson
field $\phi_a$ with the Lagrangian \cite{XGWcll}
\begin{equation}
{\cal L}_{edge} = \frac{1}{4\pi}
\partial_x \phi_a (\partial_t - \partial_x) \phi_a\,.
\label{eq:Led}
\end{equation}
The edge electron operator is given by
\begin{equation}
\psi_{edge} \propto \,:e^{-i\frac{1}{\sqrt{\nu}} \phi_a(x,t)}:
\label{eq:psiedge}
\end{equation}
while the quasiparticle operator is given by
\begin{equation}
\psi_{qp} \propto \,:e^{-i \sqrt{\nu} \phi_a(x,t)}:
\label{eq:psiqp}
\end{equation}

${\cal L}_{res}$ describes the dynamics of the electron gas
reservoir. As shown in Ref. \cite{CF}, a 2D or 3D electron gas can
be mapped to a 1D chiral Fermi liquid (FL) ($\nu = 1$) when the
tunneling is through a single point-contact. This 1D chiral Fermi
liquid is represented by a free chiral boson field $\phi_b$. ${\cal
L}_{res}$ is given by
\begin{equation}
{\cal L}_{res} = \frac{1}{4\pi} \partial_x \phi_b (\partial_t - \partial_x)
\phi_b\,.
\label{eq:Lres}
\end{equation}
In this case, the electron operator is given by
\begin{equation}
\psi_{res} \propto :e^{-i \phi_b(x,t)}:
\label{eq:el}
\end{equation}
In this paper we discuss the problem of tunneling of {\it electrons}
from the reservoir, or Fermi liquid (FL), to the FQH state and back.
Thus, some care has to be taken to keep the correct
(anti)commutation relations of the various fields. Naturally the {\it
electron operators} for the FQH state and the FL must anticommute as
they create fermion states. The bosonized formulas for the
electron operator for the edge state $\psi_{edge}$ [Eq.~(\ref{eq:psiedge})]
and  for the FL [Eq.~(\ref{eq:el})], as they stand, commute with each
other. The conventional way to fix this problem~\cite{klein} is to
multiply each operator by a suitable Klein factor (or cocycle) which
ensures the operators have the correct anticommutation properties.
The simplest choice is to define two constant
boson operators $\eta_a$
and $\eta_b$ , such that $\eta_{a,b}^\dagger\eta_{a,b} = 1$;
$[\eta_a,:e^{-i\frac{1}{\sqrt{\nu}}\phi_a(x,t)}: ]=0$
and $\eta_a^2=(-1)^{Q_b}$, where $Q_b$ is the total extra charge
at the edge described by $\phi_b$ (note that $\eta_a$ is defined in 
terms of $\phi_b$, see appendix \ref{sec:app1}).
The correct electron operators are then
$\psi_{edge} \propto \,\eta_a\; :e^{-i\frac{1}{\sqrt{\nu}}
\phi_a(x,t)}:$
and $\psi_{res} \propto \; \eta_b \;:e^{-i \phi_b(x,t)}:$ respectively.

The tunneling Lagrangian between the FQH system and the reservoir is
\begin{equation}
{\cal L}_{tun} = \Gamma \; \delta (x)\;
\eta_a^\dagger \eta_b \;
\; :e^{i[\frac{1}{\sqrt{\nu}} \phi_a(x,t) - \phi_b(x,t)]}:
+{\rm h.\ c.\ }\,,
\label{eq:Ltun}
\end{equation}
where $\Gamma$ represents the strength of the interaction and the
tunneling takes place at $x=0$ in real space. Notice, however, that the
product of the Klein factors $\eta_a^{\dagger} \eta_b$ commutes with all other
terms in the Hamiltonian and hence it is a constant of motion and,
as such, it can be absorbed in the definition of the tunneling amplitude
$\Gamma$. It is also
straightforward to show that $\eta_a^{\dagger} \eta_b$ only changes if
the total number of electrons in the combined system, edge plus
reservoir, is changed. Thus, from now on, we will drop the Klein factors
from the full Hamiltonian.

\begin{figure}
\vspace{.2cm}
\noindent
\hspace{.325 in}
\epsfxsize=2.6in
\epsfbox{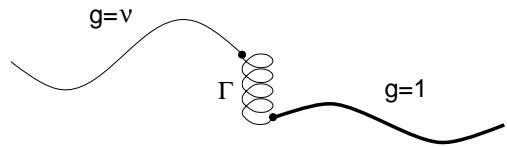}
\vspace{.5cm}
\caption{Two strings with mismatched
Luttinger parameters coupled by a tunneling interaction of strength
$\Gamma$.}
\label{fig1}
\end{figure}

By folding each chiral
boson into a semi-infinite system, one can describe the junction as
the coupling between two strings with mismatched Luttinger parameters
or compactification radii, as shown in Fig. \ref{fig1}. Such
description provides a simple way to obtain the conductance of the
junction in terms of the transmission of a pulse due to the mismatched
impedances \cite{Chklovskii&Halperin} in the case of tied strings
($\Gamma\to \infty$).

This system has two fixed points: (A) a stable fixed point at
$\Gamma=0$ (equivalent to Neumann boundary conditions because there is
no current flowing through the junction) and (B) an unstable fixed
point at $\Gamma \to \infty$ (a Dirichlet boundary condition because
the value of the field at both sides of the junction has to be the
same modulo the compactification radii). The first case
is trivial, and corresponds to two decoupled systems. The second is
solvable via a weak-strong duality symmetry, and we identify in this
case the Andreev like processes at the junction.

Since the weak coupling ($\Gamma\to 0$) fixed point is stable,
whereas the strong coupling limit ($\Gamma\to\infty$) is unstable,
an external voltage or a finite temperature sets a natural
energy scale in the problem which then separates the weak and strong
coupling regimes. It is thus meaningful to do a perturbative
expansion around {\it each} fixed point. The small parameters are
$\Gamma V^{\frac{1-\nu}{\nu}}$ near $\Gamma=0$ and $\tilde
\Gamma V^{-\frac{1-\nu}{1+\nu}}$ near $\Gamma\to\infty$, where
$\tilde\Gamma\propto\Gamma^{-\frac{1+\nu}{2\nu}}$.  The crossover at
intermediate couplings is non-perturbative but it is accessible
through the Bethe Ansatz.

Although this problem can be mapped into a free field plus
a single semi-infinite string with a local boundary action (which
belongs to a class of integrable models
\cite{Ghoshal-Zamolodchikov,fls}), the complications arise in finding
how the original quasiparticles or soliton states transform under the
map. In the soliton basis that diagonalizes the semi-infinite string
with a local boundary action, the quasiparticles (soliton states) are
scattered one by one off the point contact. That is, these
quasiparticles diagonalize the interacting Hamiltonian. However they
are not the original electrons and Laughlin quasiparticles that are
present in the asymptotic scattering states. In fact, these asymptotic
scattering states are a complicated combination of the soliton states
diagonalizing the interacting Hamiltonian. In this sense, the natural
basis for treating the interaction is not the most suitable one to
study asymptotic scattering states. In what follows we will focus on the
problem of the scattering of asymptotic quasiparticle or soliton
states at the junction.

By a suitable rotation, ${\cal L}_{tun}$ can be
written as the tunneling Lagrangian between two chiral Luttinger
liquids with an effective Luttinger parameter $g'$ as
follows\cite{CF}:
\begin{equation}
\left(\matrix{\varphi_a\cr \varphi_b}\right)
=
\left(\matrix{
\cos\theta & \sin\theta\cr
-\sin\theta & \cos\theta}\right)
\left(\matrix{
\phi_a\cr \phi_b}\right)\ ,
\label{eq:rotation}
\end{equation}
with
\begin{equation}
\cos 2\theta = \frac{2\sqrt{\nu}}{1+\nu} \,\; ,\;
\sin 2\theta = \frac{1-\nu}{1+\nu}
\label{eq:costheta}
\end{equation}
and
\begin{equation}
g'^{-1} = \frac{(1+\nu^{-1})}{2}
\label{eq:g'}
\end{equation}

Because ${\cal L}_{edge}$ and ${\cal L}_{res}$ are invariant under this O(2)
rotation, ${\cal L}$ takes the form
\begin{eqnarray}
{\cal L}& = & \frac{1}{4\pi}
\partial_x \varphi_a(\partial_t - \partial_x)\varphi_a +
\frac{1}{4\pi}
\partial_x \varphi_b(\partial_t - \partial_x)\varphi_b \nonumber\\
        &\ & \ +
\Gamma\, \delta(x)\,
e^{i \frac{1}{\sqrt{g'}}[\varphi_a(x,t) - \varphi_b(x,t)]} + h.c.
\label{eq:Lab}
\end{eqnarray}
As Eq.~(\ref{eq:Lab}) shows, the original problem involving two
different fields $\phi_a$ and $\phi_b$ with compactification radii
$R_a=\sqrt{\nu}=(2k+1)^{-1/2}$ and $R_b=1$ respectively, has been
mapped to a problem with two new fields $\varphi_a$ and $\varphi_b$
with the {\it same} compactification radius $R={\sqrt{g'}}=(k+1)^{-1/2}$ 
(see appendix \ref{sec:app2}). This
transformation mixes the states of the separate Hilbert spaces of the
decoupled systems. Naturally, the rotated states are complicated
combinations of products of states in the originally decoupled Hilbert
spaces.  Furthermore, this operation mixes states with spatial weight
even very far away from the tunnel junction.

Notice that the charge being transfered by the tunneling operator in
Eq.~(\ref{eq:Lab}) has still the value of 1 in units of the electron
charge $e$. However, the operators $e^{i\varphi_{a,b}/\sqrt{g'}}$ have
statistics $(k+1) \pi$, and thus have fermionic ($k$ even) or bosonic
($k$ odd) character. Also, notice that the natural eigenstates of the
rotated system should be viewed as solitons in terms of the original
basis.

Let us remark here that this approach can also be used to describe the
more general case of two chiral Luttinger liquids with different
filling fractions $\nu_1$ and $\nu_2$. This case corresponds to the
problem of tunneling between the edges of two different FQH systems.

\begin{figure}
\vspace{.2cm}
\noindent
\hspace{.7 in}
\epsfxsize=2.0in
\epsfbox{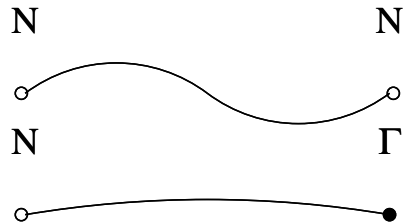}
\vspace{.5cm}
\caption{Two strings with different boundary conditions.}
\label{fig12}
\end{figure}

Once this transformation is performed the original interacting
Hamiltonian is replaced by two decoupled ones: one of a free field
$\varphi_+ =(\varphi_a + \varphi_b)/\sqrt{2} $ with a conserved current
that corresponds to the total charge of the system, and another one
$\varphi_- =(\varphi_a - \varphi_b)/\sqrt{2} $ with a backscattering
interaction. If we use the analogy with strings mentioned above, the
field $\varphi_+$ describes a string with Neumann boundary conditions
at both ends and the field $\varphi_-$ describes a string with
Neumann boundary conditions at one end and $\Gamma$ boundary
conditions at the backscattering point (see Fig.~\ref{fig12}). When
$\Gamma = 0$ this is a Neumann boundary condition and when $\Gamma
\rightarrow \infty$ it corresponds to a Dirichlet boundary
conditon. Thus, the flow from $\Gamma = 0$ to $\Gamma \rightarrow
\infty$ can be viewed as the flow from Neumann boundary conditions
to Dirichlet boundary conditions. It is on the interacting Hamiltonian for
$\varphi_-$ where the weak-strong duality transformation
\cite{Duality} is used to study the strong coupling regime.

\section{ANALYSIS AT THE FIXED POINTS}
\label{sec:analysis}
\subsection{${\bf \Gamma} = 0$ FIXED POINT}
\label{sec:gamma0}
Before analyzing the strong coupling regime, we review briefly some well known
results from the weak coupling limit $\Gamma = 0$. In this regime, the Fermi
liquid and the FQH state are decoupled from each other. If an electron
(quasiparticle) is  sent from the Fermi liquid (FQH) side of the junction it is
perfectly  reflected at the point contact. Thus there is no  net current
flowing
through the junction in any direction, corresponding to a Neumann boundary
condition at the interaction site. In this case the fields used originally to
describe electrons and quasiparticle excitations describe both the  incoming
and
the outgoing states and the probability amplitude for any scattering process
can
be calculated in a straightforward fashion. In a generic scattering process $m$
incoming electrons and $n$ incoming quasiparticles are scattered into $q$ and
$p$ electrons and quasiparticles respectively. The probability amplitude for
such a process  is proportional to:
\begin{equation}
\langle :
e^{i\sqrt{\nu}p \phi_a^{out}} e^{iq \phi_b^{out}}::
e^{-i\sqrt{\nu} n \phi_a^{in}} e^{-i m \phi_b^{in}}:
\rangle
\label{eq:m1value}
\end{equation}
where $\phi_a^{out} = \phi_a^{in}$ and $\phi_b^{out} = \phi_b^{in}$, and
the mean value is taken with respect to the quadratic action of the free
fields. Because the
fields are completely independant in this limit, this probability amplitude
can be factorized into two factors:
\begin{eqnarray}
&\ & \langle :e^{i\sqrt{\nu} p \phi_a^{out} -i \sqrt{\nu} n
\phi_a^{in}}: \rangle  \nonumber \\
&\ & \times \langle :e^{iq
\phi_b^{out}- im \phi_b^{in}}:\rangle
\label{eq:prod1}
\end{eqnarray}

The constraint imposed by charge conservation implies that $p = n$ and
$q = m$, so the number of incident electrons (quasiparticles) is the same
as the number of outgoing electrons (quasiparticles) and both charges
$Q_a$ and $Q_b$ are conserved independently. This is a perfect reflection
process as we already stated.
The set of incoming electrons and quasiparticles states can be related to
the set of
outgoing states, according to charge conservation,
by a {\it selection matrix} ${\bf M}$:
\begin{equation}
\left(\matrix{q \cr p}\right)
=
{\bf M}\left(\matrix{
m\cr n}\right)
\end{equation}
The selection matrix ${\bf M}$ transforms the incoming quantum numbers $(m,n)$
into the outgoing ones $(q,p)$.
It has the property that ${\bf M} = {\bf M}^{-1}$
which is a statement on time reversal symmetry. Since the selection matrix
relates only quantum numbers, time reversal in this language does not
involve any phase information (which is encoded in the ${\bf S}$-matrix).
Notice that in the limit $\Gamma =0$ the selection matrix ${\bf M}$ is
the identity matrix ${\bf I}$.

\subsection{${\bf \Gamma} = \infty$ FIXED POINT}
\label{sec:gammainfinity}

In the following we will focus on the strong coupling limit $\Gamma
\rightarrow \infty$. The strong coupling limit can be studied
using a weak-strong duality transformation~\cite{Duality}.
In this limit, the fields
$\varphi_{a,b}$ can be written in terms of dual fields
${\tilde\varphi}_{a,b}$ defined as
\begin{eqnarray}
\varphi_a &=&
\tilde\varphi_a \Theta(-x) + \tilde\varphi_b \Theta(x) \nonumber\\
\varphi_b &=&
\tilde\varphi_b \Theta(-x) + \tilde\varphi_a \Theta(x)\ .
\label{eq:dualfi}
\end{eqnarray}
(here $\Theta(x)$ is the step function).

The Lagrangian describing the dynamics of these fields and their interaction
is:
\begin{eqnarray}
\tilde{\cal L} & = & \frac{1}{4\pi}
\partial_x \tilde\varphi_a (\partial_t - \partial_x)\tilde\varphi_a +
\frac{1}{4\pi}
\partial_x \tilde\varphi_b (\partial_t - \partial_x)\tilde\varphi_b \nonumber\\
               & + & \tilde\Gamma \delta(x) e^{i\sqrt{g'}
[\tilde\varphi_a(x,t) - \tilde\varphi_b(x,t)]} + h.c.
\label{eq:duallag}
\end{eqnarray}
where the usual transformations $g' \rightarrow 1/g'$ and $\Gamma
\rightarrow \tilde\Gamma$ have been made.
The formulation of the problem in terms of the dual fields
${\tilde\varphi}_{a,b}$ has the advantage that, in the strong coupling
limit, these are free fields. Thus, the quasiparticles that result as
excitations of these fields are non-interacting. However they do {\it not}
describe the
original quasiparticles of the system, given in terms of the fields
$\phi_a$ (representing FQH excitations) and $\phi_b$ (representing
Fermi Liquid excitations).

By introducing the fields
\begin{equation}
{\tilde \varphi}_\pm\equiv {\frac{1}{\sqrt{2}}} ({\tilde \varphi}_a \pm {\tilde
\varphi}_b)
\label{eq:tildepm}
\end{equation}
the Lagrangian  describes two decoupled systems
\begin{eqnarray}
\tilde{\cal L}   &=& \tilde{\cal L}_+ + \tilde{\cal L}_- \nonumber \\
\tilde{\cal L}_+ &=& {\frac{1}{4\pi}}
\partial_x \tilde \varphi_+ (\partial_t - \partial_x)\tilde \varphi_+
\nonumber \\
\tilde{\cal L}_- &=& \frac{1}{4\pi} \partial_x \tilde \varphi_- (\partial_t -
\partial_x)\tilde \varphi_- \nonumber \\
                 &+&
2\tilde \Gamma \delta(x) : \cos (\sqrt{2g'} \tilde \varphi_-(x,t) ) :
\label{eq:duallag2}
\end{eqnarray}
In this representation we see that the FQH tunnel junction is equivalent to
the boundary sine-Gordon theory.

It is convenient to invert the rotation
described by Eq.~(\ref{eq:rotation}) so as to define directly free fields
$\tilde\phi_{a,b}$ dual to the original $\phi_{a,b}$:
\begin{eqnarray}
\phi_a &=& \tilde\phi_a \Theta(-x) +
\big(-\sin 2\theta\, \tilde\phi_a +
\cos 2\theta\, \tilde\phi_b\ ) \Theta(x)
\label{eq:phie}                \\
\phi_b &=&
\tilde\phi_b \Theta(-x) +
\big(\ \ \ \,\cos 2\theta\, \tilde\phi_a +
\sin 2\theta\, \tilde\phi_b\ ) \Theta(x)
\label{eq:phir}
\end{eqnarray}

\begin{figure}
\noindent
\hspace{1.0 in}
\epsfxsize=1.5in
\epsfbox{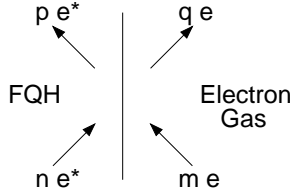}
\vspace{.5cm}
\caption{Soliton scattering process from an initial state of $n$
quasiparticles and $m$ electrons to a final state of $p$
quasiparticles and $q$ electrons.}
\label{fig2}
\end{figure}

As we stated before, the most general process consists of $m$ incoming
electrons, $n$
incoming quasiparticles, scattering into $q$ outgoing electrons and
$p$ outgoing quasiparticles ($m,n,p,q$ integers representing the {\it
net} number of electrons and quasiparticles), as shown in
Fig.~\ref{fig2}. As in the weak coupling case, the probability
amplitude of such a process is proportional to
\begin{equation}
\langle :
e^{i\sqrt{\nu}p \phi_a^{out}} e^{iq \phi_b^{out}}::
e^{-i\sqrt{\nu} n \phi_a^{in}} e^{-i m \phi_b^{in}}:
\rangle
\label{eq:mvalue}
\end{equation}
where the expectation value is taken with respect to a filled Fermi
sea of electrons and quasiparticles ($m,n,p,q$ are measured from this
level).
Using the transformation between the fields and their duals given by
Eqs.~(\ref{eq:phie},\ref{eq:phir}), Eq.~(\ref{eq:mvalue}) can be
written in terms of the dual fields, which
are free fields in the strong coupling limit. Thus Eq.~(\ref{eq:mvalue})
factorizes into a product of two free field expectation values,
\begin{eqnarray}
&\ & \langle :e^{i(-\sqrt{\nu} p \sin 2 \theta + q
\cos 2 \theta) \tilde{\phi}_a^{out} -i \sqrt{\nu} n
\tilde{\phi}_a^{in}}: \rangle  \nonumber \\
&\ & \times \langle :e^{i(\sqrt{\nu}p \cos 2\theta + q \sin 2 \theta)
\tilde{\phi}_b^{out}- im \tilde{\phi}_b^{in}}:\rangle
\label{eq:prod}
\end{eqnarray}
At strong coupling, both charges ${\tilde Q}_a$ and ${\tilde Q}_b$, are
conserved independently.
However, away from strong coupling only the
total charge ${\tilde Q}_a + {\tilde Q}_b$ is conserved; we will consider
this case shortly. For now, we focus on the infinite coupling limit.

As in the weak coupling limit, for a given set $(m,n)$ of electrons and
quasiparticles incident on
the point-contact, charge conservation imposes a constraint on the
allowed quantum numbers $(q,p)$ for the scattered states. This
constraint is written in terms of the matrix equation:
\begin{equation}
\left(\matrix{q \cr p}\right)
=
{\bf M}
\left(\matrix{
m\cr n}\right)\ ,
\qquad
{\bf M} = \,
\left(
\matrix{
\frac{1 - \nu}{1 + \nu} & \; \frac{2 \nu}{1 + \nu}  \cr
\frac{2}{1 + \nu}       & \; -\frac{1 - \nu}{1 + \nu}}
\right)\ .
\end{equation}

Besides relating incoming and outgoing quantum numbers, the selection
matrix ${\bf M}$
also contains information on the non-equilibrium
conductance of the point-contact junction.
Under the presence of an external voltage
a large number $m$ of electrons will be inciding at the junction.
The reflection and transmission coefficients for these electrons are
respectively:
\begin{eqnarray}
&R_e&=\frac{q}{m}=M_{11} \nonumber \\
&T_e&=\frac{\nu p}{m}=\nu M_{21}=1-R_e\,.
\label{eq:RT}
\end{eqnarray}

Similarly, (for a reverse applied voltage)
the coefficients for $n$ incident quasiparticles at the junction
are:
\begin{eqnarray}
&R_{qp}&=\frac{p}{n}=M_{22}  \nonumber\\
&T_{qp}&=\frac{\nu^{-1}q}{n}=\nu^{-1}M_{12}=1-R_{qp}\,.
\label{eq:RT2}
\end{eqnarray}

Notice that $T_{qp}>1$ and
$R_{qp}<0$. Such enhancement of the transmission, accompanied by a
negative reflection coefficient, is also present in N-S junctions
because of Andreev reflection. We will make this connection with
Andreev scattering more precise once we classify the soliton
scattering processes below. Finally, the conductance of the junction is
\begin{equation}
G=\frac{e^2}{h}T_e = \frac{e^2}{h} \nu T_{qp}= \frac{e^2}{h}
\frac{2\nu}{1+\nu}
\label{eq:G}
\end{equation}
in agreement with Refs. \cite{CF,Chklovskii&Halperin}
(Notice that an external voltage couples to the carriers of the system
through their charge. In
the case of incident quasiparticles, the charge of the carriers is
$\nu$, hence the factor of $\nu$ multiplying $T_{qp}$).

\section{CLASSIFICATION OF SCATTERING PROCESSESS AT ${\bf \Gamma} = \infty$ FIXED
POINT}
\label{sec:classif}

\begin{figure}
\noindent
\hspace{.65 in}
\epsfxsize=2.3in
\epsfbox{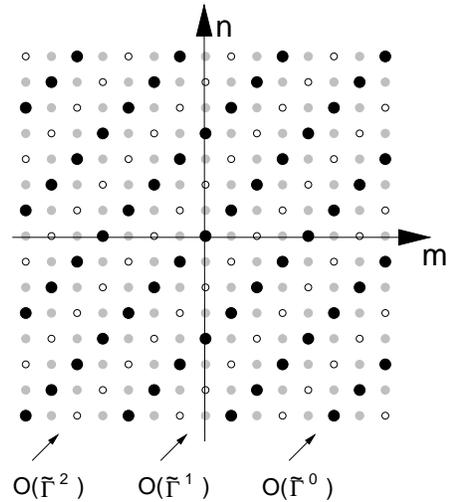}
\vspace{.5cm}
\caption{Example of the lattice of scattering processes for
 $\nu =1/7$ ($k=3$) at different orders in $\tilde\Gamma$:
black circles are zeroth order processes (strong coupling),
grey circles are first order and open circles are second order.}
\label{fig3}
\end{figure}

\begin{figure}
\noindent
\hspace{.325 in}
\epsfxsize=2.5in
\epsfbox{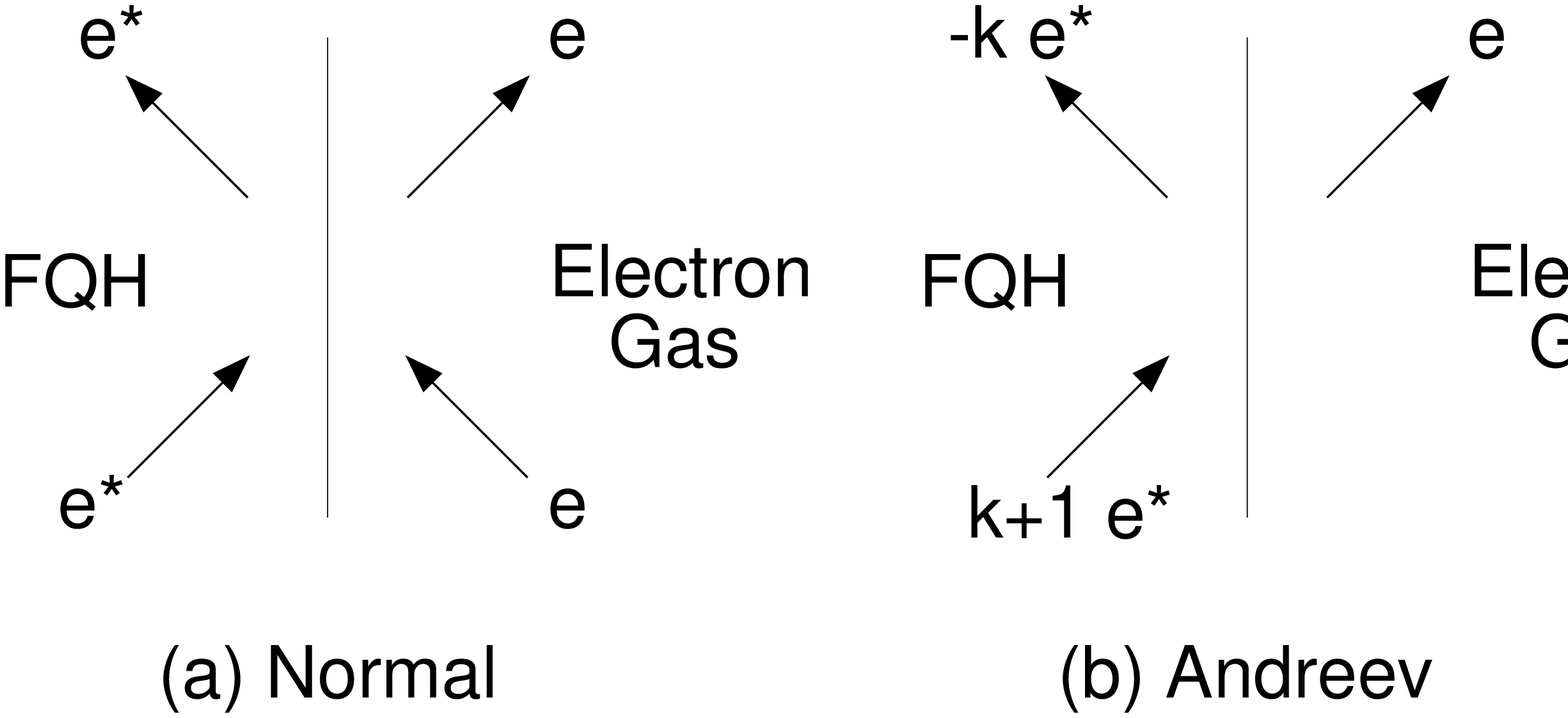}
\vspace{.5cm}
\caption{Elementary processes at interfaces between a FQH state
($\nu=\frac{1}{2k+1}$) and a normal metal at strong coupling: (a)
perfect reflection and (b) Andreev reflection with a net charge
transfer.}
\label{fig4}
\end{figure}

The requirement that $m,n,p$, and $q$ be integers constrains $(m,n)$,
as well as $(q,p)$, to
lie on a lattice (see Fig. \ref{fig3}) that is invariant under the
action of the selection matrix ${\bf M}$. Taking $\nu$ to be in the Laughlin's
sequence $\nu = 1/(2k+1)$, this lattice can be described using a
basis of two vectors
\begin{equation}
{\vec a}_1=(1,1)\,; {\vec a}_2=(0,k+1)
\label{eq:vectors}
\end{equation}
Vector ${\vec a}_1$ represents a process where one incoming electron
and one incoming quasiparticle scatter into a final state which is equal to the 
initial state (see Fig.~\ref{fig4}). As a
result there is no charge transfer between the QH system and the
reservoir through the point contact. This is the process that we will
refer to as a normal or perfect reflection.

Vector ${\vec a}_2$ represents a very different process. In this case
$k +1$ incoming quasiparticles are scattered at the point contact into
a transmitted electron to the reservoir side and $k$ reflected
quasiholes on the QH system side. Although the total charge $Q$ is
conserved, there is a net transfer of charge of one electron from the
QH system to the reservoir side. It is this process that corresponds
to Andreev reflection \cite{Andreev}, in analogy with the
normal-metal/superconductor (N-S) junction, where an incident electron
from the N side with an energy falling within the superconducting gap
is back-reflected as a hole while transfering charge $2e$ (a Cooper
pair) to the S side.  Notice that here the FQH system plays the role
of the normal metal (N), whereas the electron gas reservoir (normal
metal) plays the role of the superconductor (S).  The quasiparticles
play the role of the electrons in N and the electrons the role of the
Cooper pairs in S.  Notice also that the number of quasiparticles and
quasiholes involved in the Andreev process for FQH-N junctions depends
on the filling fraction (see Fig.~\ref{fig4}).
Our approach to the scattering of incoming and outgoing quasiparticles
and electrons at the junction can also be used to address the question
of how the properties of the electron gas reservoir are changed due
to the strong coupling to the FQH state. For example, a process like
$m=q=1, n=p=0$ where one electron is simply reflected by the junction
is not allowed in the $\Gamma\to\infty$ limit. However, it is allowed
at weak coupling, where any integer pair $(m,n)$ in the plane is
accessible. Thus, the strong coupling with a FQH state forbids some of
the scattering processes on the Fermi-liquid (FL) side of the
junction, making them an inaccessible part of the Hilbert space.
Hence, a scattering process of a state with one incoming {\it
electron} to one outgoing {\it electron} on the FL side (without
additional excitations on the FQH side) is not allowed and the
one-body sector of the ${\bf S}$-matrix vanishes.
In other words, the
overlap between an incoming state with one electron in the FL and an
outgoing state also with one electron in the FL is zero. ${\tilde
\Gamma}=0$ ($\Gamma\to\infty$) is a {\it non-Fermi liquid} fixed point.
This behavior is a manifestation of the
orthogonality catastrophe in Luttinger liquids.

In what follows, we will show that the crossover between the weak
($\Gamma =0$) and strong (${\tilde\Gamma}=0$ or $\Gamma\to\infty$) fixed
points can be understood within a simple expansion around
${\tilde\Gamma}=0$.

For small $\tilde\Gamma$ the dual fields $\tilde\phi_{a,b}$ are
coupled by the
weak perturbation ${\cal L}_{tun}=\tilde\Gamma\,\delta(x)\,
\cos \left[{ig'(\frac{1}{\sqrt{\nu} }\tilde\phi_a-\tilde\phi_b)}\right]$.
The in and out quantum numbers are now related by
\begin{equation}
\left(
\matrix{q \cr p }
\right) = {\bf M}\,
\left(
\matrix{m \cr n}
\right)\ +\ l{\bf t}\ ,
\qquad
{\bf t}=\left( \matrix{ \frac{2\nu}{1+\nu} \cr -\frac{2}{1+\nu}}
\right)
\label{eq:inoutl}
\end{equation}
and $|l|$ is the order of the expansion
($\tilde\Gamma^{|l|}$). Notice that ${\bf Mt}=-{\bf t}$, so that the
incoming and outgoing states are related by the same
Eq. (\ref{eq:inoutl}) under time reversal. The allowed values of
$(m,n)$ lie on the lattice for
$\Gamma\to\infty$ shifted by $\pm |l|$ (Fig. \ref{fig4}).
Notice that the tunneling term breaks the independent conservation of
the two charges $\tilde Q_{a,b}$, conserving only the total charge
$\tilde Q_a+\tilde Q_b$.
Hence, all integer pairs $(m,n)$ in the plane become accessible. From
this analysis we see that the strong coupling limit $\Gamma\to\infty$
can be connected to the weak coupling one $\Gamma=0$ continously.

\section{THE FQH TUNNEL JUNCTION AND QUANTUM IMPURITY PROBLEMS}
\label{sec:kondo}

In this section we will show that it is possible to establish a connection
between a Fermi liquid/FQH tunnel junction  and a two-channel Kondo problem.

The two-channel Kondo problem is a system in which
two species of band fermions (each of them in conventional Fermi
liquid states) are coupled, with strength $J$, to a spin-${\frac{1}{2}}$
magnetic impurity
localized at the origin. The renormalization group flow of the two-channel
Kondo
problem has two infrared unstable fixed points, one at $J=0$ and another at
$J=\infty$, and a non-trivial infrared stable fixed point at an intermediate
value
$J_c$ ~\cite{Nozieres}. This infrared stable fixed point controls the low
energy physics of the two-channel Kondo problem. The physics that emerges from
a study of this regime is striking. For instance, it has
been shown, first by using Bethe ansatz methods~\cite{andrei,wiegmann} and
later on by the more general approach of conformal field
theory~\cite{ludwigaffleck}, that the physics at the non-trivial fixed point
violates the Fermi liquid hypothesis. Indeed, it was found that  the phase
controlled by this fixed point is characterized by a finite, non-zero,
total entropy at zero temperature, which remains finite in the thermodynamic
limit. For the two-channel, spin-${\frac{1}{2}}$ Kondo problem, the entropy is
equal to ${\frac{1}{2}} \ln 2$.
Moreover, Affleck and Ludwig  have also shown that this is a non-Fermi liquid
fixed point in the sense that, in this regime, the one-body
${\bf S}$-matrix of the band fermions vanishes at zero frequency.
Alternatively stated, the fermion propagator no longer has a pole
but a branch cut. It has also been shown that in  the two-channel
spin-${\frac{1}{2}}$ Kondo problem~\cite{Affleck} an anisotropic exchange
coupling between the
impurity and the conduction  electrons is an irrelevant operator, but a
perturbation which induces an explicit channel anisotropy and
breaks  the degeneracy between the two channels, is a relevant operator.
The RG flows, due to the presence of the channel symmetry breaking
perturbation,  drive the  system away from the non-trivial zero-temperature
fixed point, to a strong coupling (large $J$) and large anisotropy infrared
stable fixed point. At this new fixed point the two-channel Kondo system
reduces
to two decoupled free fermion systems with different boundary conditions: one
exhibits the ordinary Kondo effect with complete screening of the impurity and
zero ground-state entropy, while the other fermion is completely decoupled from
the impurity.

The connection between the FQH junction and the two-channel Kondo problem goes
as follows. In Eq.~(\ref{eq:duallag2}) it was shown that the Lagrangian for the
FQH tunnel junction is a sum of the Lagrangians of two decoupled systems: a
free
boson (${\tilde \varphi}_+$), and a boundary sine-Gordon system (${\tilde
\varphi}_-$) with compactification radius
${\frac{1}{\sqrt{2g'}}}={\sqrt{{\frac{k+1}{2}}}}$. From the work of
Fendley, Saleur and Warner \cite{fsw} (FSW), it is known that the boundary
sine-Gordon system with this compactification radius has two fixed points: a
fixed point at
${\tilde \Gamma}=0$ and a fixed point at  ${\tilde \Gamma}=\infty$. At the
${\tilde \Gamma}=0$ fixed point, the boundary operator
$:\cos({\sqrt{2g'}}{\tilde \varphi}_-):$ is relevant (with boundary scaling
dimension ${\frac{1}{k+1}}$). This operator destabilizes the fixed point and
induces an RG flow towards the stable fixed point at  ${\tilde \Gamma}=\infty$.
FSW also found that, at the ${\tilde \Gamma}=0$ fixed point, there is a {\it
finite} ground state entropy equal to $S={\frac{1}{2}} \ln (k+1)$. At
the stable fixed point ${\tilde \Gamma}=\infty$, the ground state entropy
vanishes. Thus we see that, for the special case of $k=1$, we get the same
entropy and scaling dimensions as in the two-channel, spin-${\frac{1}{2}}$
Kondo problem. Also, in  both problems, FQH/normal metal junctions and
two-channel Kondo problem, a given perturbation drives the system from a
non-Fermi liquid fixed point to a Fermi liquid fixed point.
It is natural then to conjecture that the flow from ${\tilde
\Gamma}=0$ to ${\tilde \Gamma}=\infty$  ( or, equivalently, from
$\Gamma=\infty$ to $\Gamma=0$) can be identified with the RG trajectory in the
two-channel Kondo problem with channel anisotropy, which flows from the fixed
point at
$J_c$ in the isotropic system to the Kondo fixed point at $J=\infty$ in
the extreme anisotropic system.

\section{CONCLUSIONS}
\label{sec:conclusions}

In this paper we have discussed the physics of tunnel junctions from a
Fermi liquid to a single-edge fractional quantum Hall state.
The main focus of this work was the problem of scattering of
quasiparticles and electrons at the junction. Using the single point
contact model of the junction, introduced in Ref.~\cite{CF}, we
developed a systematic framework to classify the scattering processes
in terms of the incoming and outgoing quantum numbers for the soliton
states. We have examined the scattering processes at both the weak and
strong tunneling fixed points of the junction.

The physics of the strong coupling fixed point turned out to be quite
interesting. We have shown that, at the strong coupling fixed point,
there are selection rules that govern the scattering processes. We
have described these selection rules in terms of a selection matrix
${\bf M}$ that relates the incoming and outgoing quantum numbers of
the excitations. We found that all the scattering processes allowed in
the strong coupling limit can be viewed as a combination of two
fundamental processes: (a) normal quasiparticle-electron scattering
and (b) Andreev processes. In addition to encoding the scattering
selection rules, the elements of the matrix ${\bf M}$ give the
reflection and transmission coefficients for electrons and
quasiparticles at the junction. In particular we have shown that for
Andreev processes, which involve several quasiparticles impinging on
the junction and resulting on a transmitted electron and a number of
reflected quasiholes, there is an enhanced conductance (transmission)
and a negative reflection coefficient (on the QH side). This effect is
in complete analogy with Andreev reflection in
normal-metal/superconductor (N-S) junctions. Notice, however, that the
FQH system plays the role of N and the normal metal that of S.

We also find that the constraints imposed by charge conservation at
the strong coupling fixed point translate into forbidden processes.
For example, the one body ${\bf S}$ matrix for an incoming and
outgoing electron in the normal metal side of the junction vanishes at
strong coupling. We showed that these processes, however, become
accessible as one moves away from infinite coupling. This suggests
that the strong coupling fixed point is a non-Fermi liquid
fixed point. We then conjecture a possible connection between
FQH/normal-metal tunnel junctions and the two-channel
spin-$\frac{1}{2}$ Kondo problem. In both problems there is a flow
from a non-Fermi liquid fixed point with a finite ground state
entropy, to a Fermi liquid fixed point with a vanishing entropy.

We would like to remark here that I.\ Safi and H.\ J.\
Schulz~\cite{safi} have also discussed an analog of Andreev reflection
in tunneling processes into a Luttinger liquid with {\it attractive}
interactions. (See also the work by Furusaki and Nagaosa~\cite{furusaki},
who studied tunneling in an inhomogeneous 
Tomonaga-Luttinger liquid).
Although there is a mathematical similarity between the
Andreev processes that we discussed in this paper and the processes
studied by Safi and Schulz, they are physically quite distinct since,
in the context of the QH junctions, Andreev reflection is a
consequence of the nature of the QH edge states, which have physically
strong {\it repulsive} electron-electron interactions. Also, we would
like to point out that in the framework that we present, it is
possible to classify and describe the quasiparticle (soliton) states
which undergo Andreev reflection in addittion to identify the reflection simply
by the enhanced conductance. 

\noindent Note: While this paper was being revised, we became aware of the work
by C.\ Nayak, M.\ P.\ A.\ Fisher, A.\ W.\ W.\ Ludwig and H.\ H.\ Lin
\cite{gangoffour}. In their work, these authors have used the
approach described in this paper to describe an enhancement of the
tunneling conductance at point contact junctions of multiple Luttinger
liquid leads, in terms of Andreev reflection. Also, D.\ Maslov and 
P.\ Goldbart~\cite{gm} have discussed Andreev processes between two
adiabatically connected different non-chiral Luttinger liquids in terms of
an enhanced conductance.

\section{Acknowledgements}
\label{sec:ack}

NPS acknowledges D.L.Maslov for helpful discussions and for
pointing out the work by Safi and Schulz.
This work was supported in part by the NSF
through  grants NSF DMR94-24511 and NSF DMR-89-20538 at
the University of Illinois at Urbana-Champaign (CCC and
EF) and by the American Association of University Women (NPS).

\vspace{-.5cm}

\appendix
\section{Definition of the electron operator and Klein factors}
\label{sec:app1}

Here we consider in detail the definition of the electron operator
when there is more than one fermionic species present and the
introduction of the Klein factors needed in order to satisfy the
correct anticonmutation relations between them.

Let us consider a problem with two different types of
fermions, ${\it e.g.}$, $\psi_a$ and $\psi_b$. Because of their
fermionic character, they satisfy:
\begin{eqnarray}
\{ \psi_a(x) ; \psi_a(x')\}          &=& 0
\label{eq:one}    \\
\{ \psi_a(x) ; \psi_a^\dagger (x')\} &=& \delta(x-x')
\label{eq:two}      \\
\{ \psi_a(x) ; \psi_b(x')\}          &=& 0
\label{eq:three}
\end{eqnarray}

It is clear that the definition $\psi_{a,b} =
:e^{-\frac{i}{\sqrt{\nu}}\phi_{a,b}(x)} :$ does not satisfy
Eq.~(\ref{eq:three}) because the boson fields $\phi_a$ and $\phi_b$
commute. Thus, it is necessary to introduce new fields to obtain the
correct commutation relations. We define the electron operator
$\psi_{a,b}$ by
\begin{equation}
\psi_{a,b} = \eta_{a,b} {\cal O}_{a,b}
\label{eq:psi}
\end{equation}
where the operator ${\cal O}_{i}$ ($i=a,b$) satisfies:
\begin{eqnarray}
\{{\cal O}_i(x) , {\cal O}_i(x')\}           & =& 0
\label{eq:defO}                                           \\
\{{\cal O}_i(x) , {\cal O}_j^{\dagger}(x')\} & =& \delta_{ij} \delta(x-x')
\label{eq:defcmO}                                          \\
\bigl[ \eta_{i} , {\cal O}_{j} \bigr]       & =& 0
\label{eq:cmetaO}
\end{eqnarray}

It can be shown using Eq.~(\ref{eq:defO}, \ref{eq:cmetaO}), that Eq.~(\ref{eq:one}) is
automatically satisfied for any value of $\eta_a$. From Eq.~(\ref{eq:two}),
we see that
\begin{eqnarray}
\{ \psi_a(x) , \psi_a^{\dagger} (x')\}                             &=& \{\eta_a
{\cal O}_a(x) , {\cal O}_a^{\dagger} (x') \eta_a^{\dagger}\}
\label{eq:dagger}                                \nonumber \\
\{\eta_a {\cal O}_a(x) , {\cal O}_a^\dagger (x') \eta_a^{\dagger}\} &=& \eta_a
\eta_a^{\dagger} \delta(x-x')
\label{eq:dagger2}
\end{eqnarray}
which is satisfied if $\eta_a \eta_a^{\dagger} =\eta_a^{\dagger} \eta_a=1$. By considering that
${\cal O}_a(x) = e^{-\frac{i}{\sqrt{\nu}} \phi_a(x)}$ we propose that
\begin{equation}
\eta_a = e^{i \alpha_b p_{\phi_b}}
\label{eq:eta}
\end{equation}
where $[\phi_b(x) , p_{\phi_b}] = [\phi_{0b} , p_{\phi_b} ] = i$, that is,
$p_{\phi_b}$ is  the zero mode of $\phi_b(x)$, and $\alpha_b$ is a constant to
be determined. Notice that with the normalization chosen,
$p_{\phi_b} = \sqrt{\nu_b} Q_b$.

With this definition we can calculate Eq.~(\ref{eq:three}) as follows:
\begin{eqnarray}
\{\psi_a(x) , \psi_b(x')\} &=&0   \label{eq:long}            \\
                           &=& \{\eta_a {\cal O}_a(x) , \eta_b {\cal O}_b(x')\}
  \nonumber   \\
                           &=& e^{i \alpha_b p_{\phi_b}}
e^{-\frac{i}{\sqrt{\nu_a}} \phi_a(x)} e^{i \alpha_a p_{\phi_a}}
e^{-\frac{i}{\sqrt{\nu_b}} \phi_b(x')}
+ \nonumber     \\
                           & & e^{i \alpha_a p_{\phi_a}}
e^{-\frac{i}{\sqrt{\nu_b}} \phi_b(x')} e^{i \alpha_b p_{\phi_b}}
e^{-\frac{i}{\sqrt{\nu_a}} \phi_a(x)}
\nonumber       
\end{eqnarray}

Because of the commutation relations between $\phi(x)$ and $p_{\phi}$
\begin{equation}
e^{-\frac{i}{\sqrt{\nu_j}} \phi_j(x)} e^{i \alpha_j p_{\phi_j}} =
e^{i\frac{\alpha_j}{\sqrt{\nu_j}}} e^{i \alpha_j p_{\phi_j}}
e^{-\frac{i}{\sqrt{\nu_j}} \phi_j(x)}
\label{eq:long2}
\end{equation}

Using this result in Eq.~(\ref{eq:long}), the condition:
\begin{equation}
e^{i\frac{\alpha_a}{\sqrt{\nu_a}}} + e^{i\frac{\alpha_b}{\sqrt{\nu_b}}} = 0
\label{eq:alpha}
\end{equation}
must be satisfied. It is this condition that determines the values of
$\alpha_a$ and $\alpha_b$:
\begin{eqnarray}
\alpha_a &=& \frac{\pi}{2} \sqrt{\nu_a}  \nonumber \\
\alpha_b &=& -\frac{\pi}{2} \sqrt{\nu_b}
\label{eq:alpha2}
\end{eqnarray}

Finally, the $\eta_{a,b}$ operators are defined as:
\begin{eqnarray}
\eta_a &=& e^{i\frac{\pi}{2} Q_b} \nonumber \\
\eta_b &=& e^{-i\frac{\pi}{2} Q_a}
\label{eq:eta2}
\end{eqnarray}

Notice that the combination entering the tunneling Lagrangian in
Eq.~(\ref{eq:Ltun}) is $\eta_a^{\dagger} \eta_b =
e^{-i\frac{\pi}{2}(Q_a+Q_b)}$ and since the total charge $Q_a + Q_b$
is a constant of motion, it commutes with all the other terms in the
Hamiltonian. Also notice that $\eta^2 = e^{2i\frac{\pi}{2} Q} =
(-1)^Q$ as expected \cite{Haldane81}.

\section{Boundary Conditions and Tunneling}
\label{sec:app2}

In this appendix we give details on the rotation introduced in
Eq.~(\ref{eq:rotation}), which maps the original fields $\phi_a$ and $\phi_b$
with different compactification radii to a set of new fields $\varphi_a$
and $\varphi_b$ with the same compactification radii. We will discuss in
some detail the consistency of this rotation with the boundary
conditions, imposed by the nature of the FQH state. Here we follow the
approach first introduced by X.\ G.\ Wen~\cite{XGWcll} and discussed in
considerable detail in the review of ref.~\cite{review}.

The bosonic fields $\phi_{a,b}$ describe the fluctuations of the edges
of a given FQH state. The bulk-edge correspondence~\cite{XGWcll} implies
that, for a Laughlin FQH state with filling fraction $\nu$, the
operators $\exp(i (1/\sqrt{\nu}) \phi)$ and $\exp(i \sqrt{\nu} \phi)$
(up to Klein factors) are the operators that create electrons (with
charge $1$ and Fermi statistics) and QH quasiparticles (of charge $\nu$
and fractional statistics $\pi/\nu$). Hence, the
free boson Lagrangian and the electron and quasiparticle operators, are
{\it invariant} under the translation $\phi \rightarrow
\phi +2\pi n {\sqrt{\nu}}$, where $n$ is an integer. The invariance of
the states under these translations
has to be regarded as a symmetry of the system.
Notice that this does not imply that these are the only charged states
that can exist at the edge. In fact, since the edge has a gapless
spectrum, it is possible to construct {\it local} states with {\it
any} charge. However, these states are not globally defined since only
whole {\it electrons} can  be added or removed from the bulk FQH
fluid. Quasiparticles with fractional charge can also be created on the edges 
but at the expense of creating quasiholes in the bulk in order to maintain 
charge neutrality; {\it i.e.} by insertion of a quantum of flux in the
bulk~\cite{Laughlin}.
These facts are reflected in the theory of the edge states
through the boundary condition discussed below.

Thus, we will regard states that differ by shifts of $2\pi n R$, with
$R={\sqrt{\nu}}$ as being physically identical to each other. In other words,
the theory has been ``compactified", giving  $\phi$ the character of an
angular variable. This property dictates what are the appropriate boundary
conditions for the field $\phi$.
For an isolated FQH system, which is a droplet of perimeter $L$,
the field $\phi$ obeys periodic boundary conditions,
\begin{equation}
\phi(x) = \phi(x+L)
\label{eq:P}
\end{equation}
However, if the amount of charge in the bulk changes, 
({\it i.\ e.\/} by changing the
number of electrons or adding a quantum of flux)
$\phi$ now satisfies the more general boundary condition
\begin{equation}
\phi(x) = \phi(x+L) + 2 \pi n R
\label{eq:defR}
\end{equation}
We will follow the standard terminology in which $n$ is the
{\it winding number} of the field configuration and
$R$ the {\it compactification radius} \cite{Difrancesco}. Clearly, the boundary
condition of Eq.~(\ref{eq:defR}) is consistent with the definition of the
electron and quasiparticles operators (vertex operators).

It is worth to remark that the definition of the compactification radius
can also be interpreted as the condition for single-valuedness of the 
electron operator defined on the edges for the closed system. 
To see this let us consider the change of the 
electron operator when taken around the droplet of circumference $L$:
\begin{equation}
\psi_e(x + L) =e^{i \frac{1}{\sqrt{\nu}}\phi(x+L)} = e^{i \frac{1}{\sqrt{\nu}}\phi(x)} e^{i 2\pi n} = \psi_e(x)
\end{equation}

\noindent However the quasiparticle operator acquires an statistical phase:
\begin{equation}
\psi_{qp}(x + L) = e^{i \sqrt{\nu} \phi(x+L)} = e^{i \sqrt{\nu} \phi(x)} e^{i 2\pi n \nu} = e^{i 2\pi n \nu} \psi_{qp}(x)
\end{equation}
\noindent that is consistent with charge neutrality. As we mentioned above, in
order to have a quasiparticle in the edge of the droplet, a quasihole
must exist in the bulk. When the quasiparticle is taken around the
droplet it picks up an Aharonov-Bhom phase due to the presence of the
flux creating the quasihole in the bulk.

The winding number is related to the total charge of the edge as
follows. Let $Q$ be the {\it extra} charge at the edge due to the excitations,
namely
\begin{equation}
Q = \int_0^L dx \rho(x) = \int_0^L dx \;
{\frac{{\sqrt{\nu}}}{2\pi}}\partial_x \phi(x,t)
\label{eq:defQ}
\end{equation}
Then, consistent with this definition, we get
\begin{equation}
Q = \frac{\sqrt{\nu}}{2\pi}[\phi(x=0,t) - \phi(x=L,t)] = n\nu
\label{eq:Q}
\end{equation}
The smallest charge that can be added to the edge is the charge of one
quasiparticle (by creating a quasihole in the bulk). In this case $Q = \nu$
and $n$ counts the number of quasiparticles present in the closed system
(in this case $n=1$). 

In the FQH/normal metal junction we start with two closed systems described
by the fields $\phi_a$ with winding number $n_a$ and compactification
radius $R_a = \sqrt{\nu}$ and $\phi_b$ with winding number $n_b$
and compactification radius $R_b = 1$.
The total (extra) charge of the FQH system is $\nu n_a$
and the total (extra) charge of the normal metal system is proportional to
$n_b$.
In the presence of tunneling, there is a net charge transfer from one system
to the other but the total charge is conserved, therefore we must have
the condition
\begin{equation}
\nu n_a + n_b =0
\label{eq:neutrality}
\end{equation}
Since $n_b =1$, because the smallest amount of 
charge leaving the Fermi liquid side corresponds to one electron, we have
$n_a = \nu^{-1} = 2k+1$.

As we pointed out, the rotation defined in Eq.~(\ref{eq:rotation}), maps
the original fields to the new fields $\varphi_a$ with winding number $n'_a$
and compactification radius $R'_a$, and $\varphi_b$ with winding number
$n'_b$ and compactification radius $R'_b$. Because these new fields are  linear
combinations of the original fields, their winding numbers and compactification
radii are given by:
\begin{eqnarray}
R'_a n'_a &=&  \cos\theta R_a n_a + \sin\theta R_b n_b  \nonumber \\
R'_b n'_b &=& -\sin\theta R_a n_a + \cos\theta R_b n_b
\label{eq:rotR}
\end{eqnarray}
By replacing the expressions for $\cos\theta$, $\sin\theta$, $R_a$, $R_b$ and
the condition $\nu n_a + n_b =0$, the left hand side of Eq.~(\ref{eq:rotR}) is
given by:
\begin{eqnarray}
R'_a n'_a &=& \frac{1}{\sqrt{g'}}  \nonumber \\
R'_b n'_b &=& -\frac{1}{\sqrt{g'}} 
\label{eq:sameR}
\end{eqnarray}
In the original system we considered the presence of 
quasiparticles on the edge of the FQH side and tunneling of
electrons. In the rotated system we also consider electron tunneling
and we allow for the presence of quasiparticles on the edge.  By
choosing the compactification radius to be: $R'_{a,b}=\sqrt{g'}$ we
get $n' = 1/g'$ that is consistent with the presence of
quasiparticles on the edges of the rotated system. Notice that the
condition $g' (n'_a+n'_b) = 0$ corresponding to the conservation of 
charge in the rotated system is also automatically satisfied.


\end{multicols}


\begin{thebibliography}{99}
\bibitem{Footnote1} More precisely, the creation of a Cooper pair
is the excitation of a Goldstone mode.
\end{thebibliography}
\end{document}